\documentclass[letter]{ptptex}
\usepackage{amsmath,amssymb}

\notypesetlogo                       

\title{
The Steady State Distribution of the Master Equation
}

\author{
Mitsusada M. \textsc{Sano}
}

\inst{
Graduate School of Human and Environmental Studies, \\
Kyoto University, \\
Kyoto 606-8501, Japan
}

\abst{
The steady states of the master equation are investigated. 
We give two expressions for the steady state distribution of the master equation 
a la the Zubarev-McLennan steady state distribution, 
i.e., the exact expression and an expression near equilibrium. 
The latter expression obtained is consistent with recent attempt  
of constructing steady state theormodynamics. 
}

\begin{document}

\maketitle



In daily life, nonequilibrium steady states (NESS) 
are observed in various situations, 
such as electric current, heat conduction, and so on. 
Usually the linear response theory (Kubo formula\cite{Kubo}) 
and Onsager's reciprocal relation\cite{Onsager} 
are used to describe the NESS, i.e., the NESS near equilibrium. 
Recent advances in experimental aspects need the study of the NESS 
far from equilibrium.  
Thus understanding of the NESS is one of challenges 
in nonequilibrium statistical mechanics. 
However, our knowledge on the NESS has been limited 
until the discovery of the fluctuation theorem
\cite{ECM,ES,GC,Kurchan,LebowitzSpohn,Maes,Jarzynski,Maes_Netocny}. 
The fluctuation theorem is not restricted to near equilibrium. 
Thus the fluctuation theorem provides us with some clue 
to investigate the NESS for general settings. 

In this Letter, the NESS of the master equation is investigated.  
The master equation describes number of physical, chemical, biological, 
and even social phenomena. 
Usually the master equation was investigated by the $\Omega$-expansion
\cite{KMK,TT,TOT,TS}. 
Without the use of the $\Omega$-expansion, 
we develop a theory for the NESS of the master equation 
using the recent development of the fluctuation theorem. 
The key relation is the detailed imbalance relation 
which is used to show the fluctuation theorem. 
Thanks to the detailed imbalance relation, 
the master equation is exactly solved and 
we obtain the steady state distribution in a similar form of 
the Zubarev-McLennan steady state distribution\cite{Zubarev,McLennan}. 
However, this expression is not convenient to handle with. 
In a linear approximation near equilibrium, 
a much familiar expression is obtained, which is indeed in a form of 
the Zubarev-McLennan steady state distribution. 
Compared with the result on the NESS of the master equation
\cite{Schnakenberg,Gaspard}, 
we examine the expression obtained for the steady state distribution 
and compare it with the recent results\cite{Ruelle,Komatsu1,Komatsu2}. 


The master equation is given by 
\begin{equation}
\frac{\partial}{\partial t} P(\boldsymbol{\omega};t) = 
- \sum_{\boldsymbol{\omega}'} w_{\boldsymbol{\omega}\boldsymbol{\omega}'} 
P (\boldsymbol{\omega};t) 
+ \sum_{\boldsymbol{\omega}'} w_{\boldsymbol{\omega}'\boldsymbol{\omega}} 
P(\boldsymbol{\omega}';t)
\label{eq:master_eq}
\end{equation}
where $\boldsymbol{\omega}=(\omega_{1},\omega_{2},\dots,\omega_{N})^{t}$ is 
the discrete variable of the state. 
$P(\boldsymbol{\omega};t)$ is the probability distribution 
that the system is in state $\boldsymbol{\omega}$ at time $t$. 
$w_{\boldsymbol{\omega}\boldsymbol{\omega}'}$ is 
the transition rate that the system performs a transition 
from state $\boldsymbol{\omega}$ to state $\boldsymbol{\omega}'$ 
in a unit time. 
By definition, 
the transition rates $w_{\boldsymbol{\omega}\boldsymbol{\omega}'}$ are
non-negative. 
Equation (\ref{eq:master_eq}) can be rewritten into the following form
\cite{Schnakenberg}.
\begin{equation}
\frac{\partial}{\partial t} P(\boldsymbol{\omega};t) = 
\sum_{\boldsymbol{\omega}'} W_{\boldsymbol{\omega}'\boldsymbol{\omega}} 
P (\boldsymbol{\omega}';t), 
\label{eq:master_eq2}
\end{equation}
where
\begin{equation}
W_{\boldsymbol{\omega}'\boldsymbol{\omega}} 
= w_{\boldsymbol{\omega}'\boldsymbol{\omega}} - 
\delta_{\boldsymbol{\omega}\boldsymbol{\omega}'} 
\sum_{\boldsymbol{\omega}''}
w_{\boldsymbol{\omega}\boldsymbol{\omega}''}.
\label{eq:W}
\end{equation}
Since the master equation conserves the total probability,   
the transition rates $W_{\boldsymbol{\omega}\boldsymbol{\omega}'}$ 
satisfy the following condition.
\begin{equation}
\sum_{\boldsymbol{\omega}'} W_{\boldsymbol{\omega}\boldsymbol{\omega}'} = 0.
\label{eq:constraint}
\end{equation}
This relation can be confirmed directly. 
Therefore, an alternative form of the master equation is obtained.
\begin{equation}
\frac{\partial}{\partial t} P(\boldsymbol{\omega};t) = 
- \sum_{\boldsymbol{\omega}'} W_{\boldsymbol{\omega}\boldsymbol{\omega}'} 
P (\boldsymbol{\omega};t) 
+ \sum_{\boldsymbol{\omega}'} W_{\boldsymbol{\omega}'\boldsymbol{\omega}} 
P(\boldsymbol{\omega}';t)
\label{eq:master_eq3}
\end{equation}
Hereafter we consider this master equation. 
Note that the transition rates $W_{\boldsymbol{\omega}\boldsymbol{\omega}'}$ 
are no longer non-negative and the diagonal elements 
$W_{\boldsymbol{\omega}\boldsymbol{\omega}}$ are non-positive. 

Now we assume the detailed imbalance relation
\cite{ECM,ES,GC,Kurchan,LebowitzSpohn,Maes,Jarzynski,Maes_Netocny}
\footnote{Sometimes it is called 
the nonequilibrium detailed balance relation.}:  
\begin{equation}
\frac{P(\boldsymbol{\omega};t-0)W_{\boldsymbol{\omega}\boldsymbol{\omega}'}}
{P(\boldsymbol{\omega}';t+0)W_{\boldsymbol{\omega}'\boldsymbol{\omega}}}
=
\exp \left [ \sigma_{\boldsymbol{\omega}\boldsymbol{\omega}'}(t)\right ],
\label{eq:DIR}
\end{equation}
where $\sigma_{\boldsymbol{\omega}\boldsymbol{\omega}'}(t)$ is 
the entropy production for one jump 
$\boldsymbol{\omega} \rightarrow \boldsymbol{\omega}'$. 
Using Eq.~(\ref{eq:DIR}), Eq.~(\ref{eq:master_eq3}) is rewritten as
\begin{equation}
\frac{\partial}{\partial t} P(\boldsymbol{\omega};t) = 
P (\boldsymbol{\omega};t) 
\sum_{\boldsymbol{\omega}'} 
\left ( \exp[-\sigma_{\boldsymbol{\omega}\boldsymbol{\omega}'}(t)] - 1
\right ) W_{\boldsymbol{\omega}\boldsymbol{\omega}'}. 
\label{eq:master_eq_2}
\end{equation}
Equation (\ref{eq:master_eq_2}) is easily solved. 
\begin{equation}
P(\boldsymbol{\omega};t) = 
C(\boldsymbol{\omega};0) \exp \left [ 
\int_{0}^{t} \; dt' \; \sum_{\boldsymbol{\omega}'} 
\left ( \exp[-\sigma_{\boldsymbol{\omega}\boldsymbol{\omega}'}(t')] - 1
\right ) W_{\boldsymbol{\omega}\boldsymbol{\omega}'}
\right ],
\label{eq:pdf_1}
\end{equation}
where $C(\boldsymbol{\omega};0)$ will be determined later. 
As in the standard definition, here we set 
\begin{equation}
P(\boldsymbol{\omega};t) = \exp [ -S(\boldsymbol{\omega};t) ].  
\label{eq:pdf_0}
\end{equation}
Equation (\ref{eq:pdf_1}) is rewritten as
\begin{eqnarray}
P(\boldsymbol{\omega};t) & = &  
C(\boldsymbol{\omega};0) \exp \left [ 
\int_{0}^{t} \; dt' \; 
\frac{\partial P(\boldsymbol{\omega};t')/\partial t'}{
P(\boldsymbol{\omega};t')} \right ] \nonumber \\
& = & C(\boldsymbol{\omega};0) \exp \left [ 
-\int_{0}^{t} \; dt' \; 
\dot{S}(\boldsymbol{\omega};t') \right ] \nonumber \\
& = & C(\boldsymbol{\omega};0) \exp \left [ 
S(\boldsymbol{\omega};0)-S(\boldsymbol{\omega};t)
 \right ] 
\end{eqnarray}
To be consistent with Eq.~(\ref{eq:pdf_0}), 
it should be $C(\boldsymbol{\omega};0)=\exp[-S(\boldsymbol{\omega};0)]$. 
Thus, Eq.~(\ref{eq:pdf_1}) is a formal (exact) solution. 

For the NESS, the steady state distribution is given by 
\begin{eqnarray}
P^{st}(\boldsymbol{\omega}) & = & 
\exp \left [ -S(\boldsymbol{\omega};0) + 
\int_{0}^{\infty} \; dt' \; \sum_{\boldsymbol{\omega}'} 
\left ( \exp[-\sigma_{\boldsymbol{\omega}\boldsymbol{\omega}'}(t')] - 1
\right )W_{\boldsymbol{\omega}\boldsymbol{\omega}'}
\right ] \nonumber \\
& = & 
\exp \left [ -S(\boldsymbol{\omega};0) + 
\int_{0}^{\infty} \; dt' \; \sum_{\boldsymbol{\omega}'} 
W_{\boldsymbol{\omega}\boldsymbol{\omega}'}
\sum_{n=1}^{\infty}\frac{1}{n!}
(-\sigma_{\boldsymbol{\omega}\boldsymbol{\omega}'}(t'))^{n}
\right ].
\label{eq:ss_1}
\end{eqnarray}
This is the first main result 
and is nothing but the exact expression of the NESS, 
which is similar to the Zubarev-McLennan steady state distribution. 
However, Eq.~(\ref{eq:ss_1}) is not in a useful form. 
Thus, we consider a NESS near equilibrium. 
Near equilibrium, the entropy production is small. 
So we can approximate as $\exp(-\sigma) \approx 1 - \sigma$, 
i.e., the linear approximation near equilibrium. 
Then, we obtain the probability distribution 
\begin{equation}
P(\boldsymbol{\omega};t) \simeq
\exp \left [ -S(\boldsymbol{\omega};0) 
- \int_{0}^{t} \; dt' \; \sum_{\boldsymbol{\omega}'} 
\sigma_{\boldsymbol{\omega}\boldsymbol{\omega}'}(t') 
W_{\boldsymbol{\omega} \boldsymbol{\omega}'}
\right ].
\label{eq:pdf_2}
\end{equation}
Near equilibrium, the steady state distribution is approximated as 
\begin{eqnarray}
P^{st}(\boldsymbol{\omega}) & \simeq & 
\exp \left [ -S(\boldsymbol{\omega};0)  -
\int_{0}^{\infty} \; dt' \; \sum_{\boldsymbol{\omega}'} 
\sigma_{\boldsymbol{\omega} \boldsymbol{\omega}'}(t')
W_{\boldsymbol{\omega} \boldsymbol{\omega}'}
\right ] \nonumber \\
& = & 
\exp \left [ -S(\boldsymbol{\omega};0)  
-\int_{0}^{\infty} \; dt' \; 
\{  
\Sigma(\boldsymbol{\omega};t') - 
\left \langle J(\boldsymbol{\omega} )\right \rangle 
\} 
\right ],
\label{eq:ss_2}
\end{eqnarray}
where
\begin{equation}
\Sigma(\boldsymbol{\omega};t)  =  \sum_{\boldsymbol{\omega}'}
\{ S(\boldsymbol{\omega}';t)-S(\boldsymbol{\omega};t) \} 
W_{\boldsymbol{\omega}\boldsymbol{\omega}'}, 
\end{equation}
and
\begin{equation}
\left \langle J(\boldsymbol{\omega} )\right \rangle   =  
- \sum_{\boldsymbol{\omega}'} 
W_{\boldsymbol{\omega}\boldsymbol{\omega}'} 
\ln \frac{W_{\boldsymbol{\omega}\boldsymbol{\omega}'}}
{W_{\boldsymbol{\omega}'\boldsymbol{\omega}}}.
\end{equation}
This is the second main result. 
$\Sigma(\boldsymbol{\omega};t)$ is 
the total entropy production including the incoming entropy production flow, 
i.e.,  $\frac{1}{\tau}\mathit{\Delta}S(\boldsymbol{\omega})$. 
$\left \langle J(\boldsymbol{\omega} )\right \rangle $ is 
the entropy production flow, 
i.e., $\frac{1}{\tau}\mathit{\Delta}_{e}S(\boldsymbol{\omega})$. 
Thus, 
$\Sigma(\boldsymbol{\omega};t)
-\left \langle J(\boldsymbol{\omega} )\right \rangle$ is 
the (internal) entropy production, i.e., 
$\frac{1}{\tau}\mathit{\Delta}_{i}S(\boldsymbol{\omega}) = 
\frac{1}{\tau}(\mathit{\Delta}S(\boldsymbol{\omega}) 
- \mathit{\Delta}_{e}S(\boldsymbol{\omega}))$. 
Thus, the argument of the exponential function in the second line 
of Eq.~(\ref{eq:ss_2}) 
expresses the time integration of the minus of the excess entropy production 
or the internal entropy production. 

The averaged entropy production in the NESS is given by\cite{Schnakenberg,Gaspard} 
\begin{eqnarray}
\left \langle  \sigma \right \rangle & = & 
\sum_{\boldsymbol{\omega},\boldsymbol{\omega}'} 
P^{st}(\boldsymbol{\omega})W_{\boldsymbol{\omega}\boldsymbol{\omega}'} 
\ln
\frac{P^{st}(\boldsymbol{\omega})W_{\boldsymbol{\omega}\boldsymbol{\omega}'} }
{P^{st}(\boldsymbol{\omega}')W_{\boldsymbol{\omega}'\boldsymbol{\omega}} }. 
\label{eq:av_ep}
\end{eqnarray}
This expression can be written in terms of Kolmogorov-Sinai (KS) entropy
\cite{Gaspard}. 
\begin{eqnarray}
\left \langle  \sigma \right \rangle & = & h^{R}- h,
\end{eqnarray}
where $h$ is KS entropy and 
$h^{R}$ is KS entropy for the reversed process. 
Since 
\begin{equation}
\sum_{\boldsymbol{\omega}'} 
\sigma_{\boldsymbol{\omega}\boldsymbol{\omega}'}(t) 
W_{\boldsymbol{\omega}\boldsymbol{\omega}'} = 
\sum_{\boldsymbol{\omega}'} 
W_{\boldsymbol{\omega}\boldsymbol{\omega}'} 
\ln \frac{P(\boldsymbol{\omega};t)W_{\boldsymbol{\omega}\boldsymbol{\omega}'}}
{P(\boldsymbol{\omega}';t)W_{\boldsymbol{\omega}'\boldsymbol{\omega}}}, 
\end{equation}
thus this quantity resembles the content inside of the sum 
in the right hand side of Eq.~(\ref{eq:av_ep}). 
This quantity can be identified with the excess entropy production 
in state $\boldsymbol{\omega}$. 

Consider the entropy in state $\boldsymbol{\omega}$. 
From Eq.~(\ref{eq:pdf_2}), 
we obtain
\begin{eqnarray}
S(\boldsymbol{\omega};t) & = & -\ln P(\boldsymbol{\omega};t)\nonumber \\
& = &  
S(\boldsymbol{\omega};0) + \nonumber \\
& & 
\sum_{\boldsymbol{\omega}'} 
\int_{0}^{t} dt' 
\left \{ W_{\boldsymbol{\omega}\boldsymbol{\omega}'} 
( S(\boldsymbol{\omega}';t') - S(\boldsymbol{\omega};t') ) + 
W_{\boldsymbol{\omega}\boldsymbol{\omega}'} 
\ln \frac{W_{\boldsymbol{\omega}\boldsymbol{\omega}'}}
{W_{\boldsymbol{\omega}'\boldsymbol{\omega}}} \right \}. 
\label{eq:S}
\end{eqnarray}
This equation expresses the time-evolution 
of the entropy $S(\boldsymbol{\omega};t)$. 
Taking time derivative of Eq.~(\ref{eq:S}) and 
noting Eq.~(\ref{eq:constraint}), 
we have
\begin{equation}
\frac{d}{dt}S(\boldsymbol{\omega};t) = 
\sum_{\boldsymbol{\omega}'} 
W_{\boldsymbol{\omega}\boldsymbol{\omega}'}S(\boldsymbol{\omega}';t) 
- \left \langle J(\boldsymbol{\omega}) \right \rangle. 
\end{equation}
If we use the matrix notation for the transition rates 
$W_{\boldsymbol{\omega}\boldsymbol{\omega}'}$, 
and the vector notations for 
the entropy $S(\boldsymbol{\omega};t)$ and the steady entropy production current 
$\left \langle J(\boldsymbol{\omega}) \right \rangle$, 
we have 
\begin{equation}
\dot{\boldsymbol{S}}(t) = {\sf W} \boldsymbol{S}(t) - \boldsymbol{J}.
\end{equation}
This can be easily solved as
\begin{equation}
\boldsymbol{S}(t) = 
e^{{\sf W}t} \boldsymbol{S}(0) - \int_{0}^{t}ds\;  
e^{-{\sf W}(s-t)}\boldsymbol{J}.
\end{equation}
In the NESS, the following condition is satisfied.
\begin{equation}
\sum_{\boldsymbol{\omega}'} 
W_{\boldsymbol{\omega}'\boldsymbol{\omega}}P^{st}(\boldsymbol{\omega}') = 0. 
\label{eq:eigen_prob}
\end{equation}
Here we have set the right hand side of Eq.~(\ref{eq:master_eq2}) to be zero. 
Eq.~(\ref{eq:eigen_prob}) implies 
that there exist zero eigenvalues for the matrix ${\sf W}$. 
Now we assume that at $t=\infty$, the state reaches the NESS. 
As a result, the steady state distribution is given by 
$P^{st}(\boldsymbol{\omega})  =  \exp [-S^{st}(\boldsymbol{\omega})] 
 =  \exp [-S(\boldsymbol{\omega};\infty)]$. Thus, we have 
\begin{eqnarray}
& & P^{st}(\boldsymbol{\omega}) \nonumber \\
& \simeq & 
\exp \left [ - 
\lim_{t\rightarrow \infty} 
\sum_{\boldsymbol{\omega}'}
\{
(e^{{\sf W}t})_{\boldsymbol{\omega}\boldsymbol{\omega}'} 
S(\boldsymbol{\omega}';0)
- 
\int_{0}^{t} ds\; 
(e^{-{\sf W}(s-t)})_{\boldsymbol{\omega}\boldsymbol{\omega}'} 
J_{\boldsymbol{\omega}'}
\}\right ].
\label{eq:ss_6}
\end{eqnarray}
where
\begin{equation}
J_{\boldsymbol{\omega}'}=
-
\sum_{\boldsymbol{\omega}''}
W_{\boldsymbol{\omega}'\boldsymbol{\omega}''} 
\ln \frac{W_{\boldsymbol{\omega}'\boldsymbol{\omega}''} }
{W_{\boldsymbol{\omega}''\boldsymbol{\omega}'} }.
\label{eq:J_omega_d}
\end{equation}
This is the third main result. 
The rank of the matrix ${\sf W}$ is smaller than its matrix size. 
We assume that the matrix ${\sf W}$ has only one zero eigenvalue 
and the others are negative eigenvalues, 
since the diagonal elements of the matrix ${\sf W}$ have 
negative values (See Eq.~(\ref{eq:W})). 
The first term in the argument of the right hand side of Eq.~(\ref{eq:ss_6}) 
converges to the element of the eigenvector 
corresponding to the zero eingevalue, i.e., the NESS. 



We have demonstrated that near equilibrium, 
the solution of the master equation is solved analytically.  
The time-evolution of the entropy and the distribution function was obtained  
and the steady state distribution near equilibrium was evaluated. 
To include nonlinear effects far from equilibrium, 
i.e., beyond the linear approximation, 
one should include the nonlinear terms ($n=2,3,\cdots$) 
in the argument of the exponential function 
in the second line of Eq.~(\ref{eq:ss_1}). 
But this task would be tedious. 
At present, there is no results in this direction. 

The expression of Eq.~(\ref{eq:ss_2}) is very similar to the result of 
recent attempt of constructing steady state 
thermodynamics\cite{Ruelle,Komatsu1,Komatsu2}. 
Now we rewrite Eq.~(\ref{eq:ss_2}) into the form of the expression 
derived by Komatsu and Nakagawa\cite{Komatsu1}. 
Noting that $\sigma_{\boldsymbol{\omega}\boldsymbol{\omega}'}(t) = 
-\sigma_{\boldsymbol{\omega}'\boldsymbol{\omega}}(t)$,
Eq.~(\ref{eq:ss_2}) can be rewritten as 
\begin{equation}
P^{st}(\boldsymbol{\omega}) \simeq 
\exp \left [ -S(\boldsymbol{\omega};0) + 
\frac{1}{2}\int_{-\infty}^{0} \; dt' \; \sum_{\boldsymbol{\omega}'} 
\sigma_{\boldsymbol{\omega}' \boldsymbol{\omega}}(t')
W_{\boldsymbol{\omega} \boldsymbol{\omega}'}
-
\frac{1}{2}\int_{0}^{\infty} \; dt' \; \sum_{\boldsymbol{\omega}'} 
\sigma_{\boldsymbol{\omega} \boldsymbol{\omega}'}(t')
W_{\boldsymbol{\omega} \boldsymbol{\omega}'}
\right ].
\label{eq:ss_4}
\end{equation}
This equation is similar to Eq.~(15a) and (15b) in Ref.~\citen{Komatsu1}. 

Finally one question remains: 
"{\it Is the Zubarev-McLennan steady state distribution 
for the NESS near equilibrium?}" 
Eq.~(\ref{eq:ss_2}) (i.e., the expression near equilibrium) seems to correspond 
to the Zubarev-McLennan steady state distribution\cite{Zubarev,McLennan}. 

The author is grateful to K.~Kitahara and H.~Tomita 
for continuous encouragement and 
to S.~Takesue for pointing out a mistake.

%
%



\end{document}